\documentclass[11pt,a4paper]{article}
\usepackage[margin=2.6cm]{geometry}
\usepackage{amsmath,amssymb}

\usepackage{authblk}

\usepackage{graphicx}
\usepackage{booktabs}
\usepackage{xcolor}
\definecolor{linkblue}{RGB}{25,60,120}
\usepackage[colorlinks=true,linkcolor=linkblue,citecolor=linkblue,urlcolor=linkblue]{hyperref}
\usepackage{microtype}

\newcommand{\Msun}{M_\odot}
\newcommand{\kB}{k_{\rm B}}
\newcommand{\Mdot}{\dot{M}}

\begin{document}
	
\title{How greedy is the Universe?\\ An entropy ledger and a causal envelope\\ for early black hole growth}

\author[1]{Fernando Izaurieta\thanks{
		\href{mailto:fernando.izaurieta@uss.cl}
		{\texttt{fernando.izaurieta@uss.cl}}}}

\author[1]{Cristian Quinzacara\thanks{
		\href{mailto:cristian.quinzacara@uss.cl}
		{\texttt{cristian.quinzacara@uss.cl}}}}	

\author[2,3]{Omar Valdivia\thanks{
		\href{mailto:ovaldivi@unap.cl}
		{\texttt{ovaldivi@unap.cl}}}}	

\affil[1]{Departamento de Ciencias Exactas, Facultad de Ingeniería, Universidad San Sebastián, Concepción, Chile}

\affil[2]{Instituto de Ciencias Exactas y Naturales, Universidad Arturo Prat, Playa Brava 3256, 1111346, Iquique, Chile}

\affil[3]{Facultad de Ciencias, Universidad Arturo Prat, Avenida Arturo Prat Chacón 2120, 1110939, Iquique, Chile}

\date{}

\maketitle

\begin{abstract}
	\noindent
	The entropy of the observable Universe is dominated, by fifteen orders of magnitude, by the horizons of supermassive black holes, and gravitational collapse into black holes has often been proposed as the Universe's preferred channel of entropy production. We examine this suggestion quantitatively, and the analysis largely refutes it. Three results are presented. First, an entropy production history: the rate $dS/dt(z)$ resolved by channel, from the star formation history and a Soltan-normalized accretion history. Black hole horizon growth exceeds all radiative channels by a factor $\sim 10^{16-17}$ after the first seeds; the total rate peaked near $z\approx 1$--$2$ and has since declined; the late peak is structural, because horizon entropy production weights accretion by the accretor's mass and the most massive accretors are assembled last. Second, a causal-hydrodynamic envelope: the fastest entropy-producing trajectory permitted by causality and gas dynamics, $\dot M \sim c_s^3/G$ in atomic-cooling halos, which reaches $10^{7\pm1}\,M_\odot$ by $z\approx 8$--$10$ without radiative throttling, and whose phenomenology (obscured, radiatively inefficient, X-ray weak, red) closely resembles the JWST little red dots. The realized-to-envelope ratio defines an efficiency $F(z)$, which remains between $10^{-7}$ and $10^{-4}$ throughout $4\le z\le 12$, even under the most generous JWST-era bracket. Third, order-of-magnitude no-go results: Boltzmann weighting of collapse channels by entropy gain, dissipative adaptation applied to self-gravitating systems, and any strong maximum-entropy-production principle in cosmology all fail at leading order, by factors of up to $10^{16}$. The Universe produces entropy overwhelmingly through black holes, yet remains far below its own envelope at all times. Whatever selects cosmic structure, it does not maximize entropy production.
\end{abstract}
\vspace{2mm}

\section{Introduction}
\label{sec:intro}

By the time the Universe was one second old, causality already permitted black holes of two hundred thousand solar masses: the horizon mass at that epoch is $c^3 t/G \approx 2\times10^{5}\,\Msun$. Had the baryons of the observable Universe collapsed into such horizons at that time, the total entropy would have been $1.6\times10^{105}\,\kB$, five times what the Universe has actually accumulated over $13.8$ Gyr~\cite{EganLineweaver2010,Profumo2024}. The second law permitted it, and causality permitted it, yet it did not occur. The Universe, smooth to one part in $10^{5}$, did not take this path and has since recovered only a small fraction of the entropy it thereby left unrealized.

This paper treats that refusal as a quantitative datum. The underlying temptation is both old and well established. Since Frautschi's analysis of entropy in an expanding, causally limited universe~\cite{Frautschi1982}, and through subsequent work by Lineweaver and collaborators~\cite{LineweaverEgan2008,Lineweaver2014,PatelLineweaver2017}, it has been recognized that (i) the entropy budget of the observable Universe is overwhelmingly dominated by supermassive black hole (SMBH) horizons, $S_{\rm SMBH}\sim10^{104}\,\kB$ against $\sim10^{89}\,\kB$ in photons and neutrinos~\cite{EganLineweaver2010,Profumo2024}, and (ii) an expanding universe moves \emph{away} from equilibrium because the maximum entropy available to it grows faster than the entropy actually realized~\cite{Frautschi1982,Layzer1975}. From these observations it is natural, and increasingly common in varying forms, to suggest that gravitational collapse into black holes is the process the Universe preferentially realizes: that structure formation, and in particular the unexpectedly early assembly of SMBHs now documented by JWST \cite{Wang2021,Maiolino2024,LRDreview2026}, might admit a thermodynamic explanation, perhaps through some maximum-entropy-production principle (MEPP)~\cite{Lineweaver2014,MartyushevSeleznev2006,Martyushev2021}.

The purpose of this paper is to replace that metaphysical suggestion with three measurable quantities and two quantitative refutations. We state the conclusion immediately because it runs counter to the suggestion: once one computes how much entropy the Universe produces, through which channels, and how much it \emph{could} produce within the limits imposed by causality and gas dynamics, the result is not a Universe that maximizes entropy production. Rather, it is a Universe that forgoes a factor of $\sim 3\times10^{16}$ whenever a gas cloud fragments into stars instead of collapsing into a black hole, as occurs in nearly every case; that remains four to seven orders of magnitude below its own entropic envelope at every epoch; and whose most effective entropy-producing channel, early monolithic collapse, requires \emph{suppressing} local dissipation rather than maximizing it. Structure formation therefore behaves like a greedy algorithm: it optimizes locally, step by step, without access to the global maximum\footnote{``Greedy'' is used here in the algorithmic sense, without moral judgment, although the Universe's performance remains suggestive. If the tendency to produce entropy through black holes is strong, it is nevertheless realized only weakly.}.

We present three results that are, to the best of our knowledge, new. \emph{First}, the entropy production \emph{history} of the Universe, $dS/dt(z)$ resolved by channel (Sec.~\ref{sec:ledger}, Fig.~\ref{fig:ledger}). The budget at the present epoch is known~\cite{EganLineweaver2010,Profumo2024}; its rate history, with the black hole channel normalized through the Soltan argument and the radiative channels through the star formation history, does not appear to have been assembled previously. Two features deserve emphasis: black hole horizon growth exceeds every radiative channel by sixteen to seventeen orders of magnitude at all epochs after the first seeds, and the total production rate peaked near $z\approx1$--$2$ and has declined since; the Universe's entropy production has passed its maximum. \emph{Second}, the definition and computation of a \emph{causal-hydrodynamic envelope} for black hole growth, together with the dimensionless efficiency $F(z)$, defined as the ratio of realized to envelope entropy production (Secs.~\ref{sec:envelope} and \ref{sec:efficiency}, Figs.~\ref{fig:envelope} and \ref{fig:efficiency}). \emph{Third}, a set of order-of-magnitude no-go results for entropic teleologies in structure formation (Sec.~\ref{sec:nogo}), which are robust in a way that the present observational situation is not: they do not depend on any contested JWST mass estimate. Section~\ref{sec:physics} then develops the physical interpretation of all three results; once their structure is made explicit, none is obscure.

A fourth point is an observation rather than a result. The envelope trajectory, monolithic collapse at the atomic-cooling rate with radiative feedback suppressed, is not merely abstract: it describes an object that grows while obscured, radiates inefficiently, is X-ray weak, and presents a red photosphere. This closely matches the phenomenology of the JWST little red dots \cite{LRDreview2026,Greene2024,BHstar2025}. Section~\ref{sec:LRD} presents this relation carefully as consistency rather than derivation, and identifies what the ongoing revision of the relevant mass estimates \cite{UHZ1reanalysis2026,Overmassive2026} can and cannot alter in the argument.

Two qualifications frame the analysis that follows. This paper introduces no new dynamics and ``predicts'' nothing beyond what the underlying dynamics already predicts; its contribution is a bookkeeping framework in which extremal claims become testable and, in most cases, fail. It also takes no position on whether a refined, constraint-qualified extremal principle might survive; it requires only that the relevant constraints be specified in advance, because a principle whose constraints are adjusted after each observation has no independent explanatory content.

Throughout we use Planck 2018 parameters, work with comoving densities, and count horizon entropy in the Bekenstein--Hawking form $S=4\pi G \kB M^2/\hbar c$, so that $S(M) = 1.05\times10^{77}\,(M/\Msun)^2\,\kB$. The cosmic event horizon is maintained in a separate ledger for the reasons discussed in Sec.~\ref{sec:discussion}.

\section{The entropy ledger}
\label{sec:ledger}

\subsection{Channels}
\label{sec:channels}

A black hole of mass $M$ accreting at a rate $\Mdot$ increases its horizon entropy at
\begin{equation}
	\frac{dS_{\rm BH}}{dt} \;=\; \frac{8\pi G \kB}{\hbar c}\, M \Mdot
	\;=\; 2.1\times10^{85}
	\left(\frac{M}{10^{8}\,\Msun}\right)
	\left(\frac{\Mdot}{\Msun\,{\rm yr}^{-1}}\right) \kB\,{\rm yr}^{-1}.
	\label{eq:dSdtBH}
\end{equation}
The linear weighting by $M$ implies that cosmic entropy production is dominated by the most massive black holes that are actively accreting, a point that will become important below. By comparison, the entropy carried by the radiation emitted during the same accretion episode is smaller than the horizon increase by
\begin{equation}
	\frac{\dot S_{\rm rad}}{\dot S_{\rm hor}} \;\sim\; \frac{\epsilon}{1-\epsilon}\,
	\frac{T_{\rm H}}{T_{\rm disc}},
	\label{eq:radhor}
\end{equation}
where $\epsilon$ is the radiative efficiency of accretion: the fraction $\epsilon$ of the accreted rest-mass energy is radiated away, while the remaining $1-\epsilon$ crosses the horizon.

Thus, for representative values $\epsilon=0.1$, $T_{\rm disc}\sim3\times10^{4}\,\mathrm{K}$ for the disk photosphere, and $T_{\rm H}(10^8 \Msun)=6\times10^{-16}\,\mathrm{K}$, one obtains
\begin{equation}
	\frac{\dot S_{\rm rad}}{\dot S_{\rm hor}} \approx 2\times10^{-21}.
	\label{eq:radhor2}
\end{equation}
A quasar, despite being among the most conspicuous dissipative structures in the sky, therefore contributes only marginally through its radiation; the dominant entropy increase occurs at the horizon.

The principal radiative channel of the Universe is starlight, much of which is reprocessed by dust into the far infrared. If a comoving volume forms stars at a rate $\psi(z)$ and promptly radiates a fraction $\epsilon_\star$ of the newly formed rest mass, the photon entropy production is
\begin{equation}
	\frac{dS_\gamma}{dt} \;=\; \frac{4}{3}\,\epsilon_\star\, \psi\, c^2
	\left[\frac{f_{\rm IR}}{\kB T_{\rm dust}} + \frac{1-f_{\rm IR}}{\kB T_\star}\right],
	\label{eq:dSdtgamma}
\end{equation}
with $T_{\rm dust}\approx35$ K, $T_\star\approx5800$ K, and $f_{\rm IR}\approx0.5$--$0.8$ denoting the dust-reprocessed fraction. Neutrinos from stellar collapse and nuclear burning contribute at a still smaller level and are omitted from the figures.\footnote{Core-collapse neutrinos carry $\sim3\times10^{53}$ erg per event at $T\sim4\times10^{10}$ K, which works out to $\sim5\times10^{56}\,\kB$ per solar mass of stars formed: five orders below the dust channel per unit star formation, eleven below Eq.~(\ref{eq:dSdtBH}) per unit accretion.}

\subsection{Inputs}
\label{sec:inputs}

For $\psi(z)$ we use the Madau--Dickinson fit \cite{MadauDickinson2014},
\begin{equation*}
	\psi(z)=0.015\,\frac{(1+z)^{2.7}}{1+\left(\dfrac{1+z}{2.9}\right)^{5.6}}\;\Msun\,{\rm yr}^{-1}\,{\rm Mpc}^{-3},
\end{equation*}
with $\epsilon_\star=7\times10^{-4}$, which reproduces the local bolometric luminosity density. For the black hole accretion rate density $\psi_{\rm BH}(z)$, we adopt the same functional form and normalize it through the Soltan argument \cite{Soltan1982}: the time integral of $\psi_{\rm BH}$ must equal the local SMBH mass density, taken as $\rho_{\rm BH,0}=4.2^{+1.8}_{-1.7}\times10^{5}\,\Msun\,{\rm Mpc}^{-3}$ \cite{Shankar2009}. This form agrees with determinations based on the bolometric quasar luminosity function \cite{Shen2020} to within a factor of order two near the peak, a difference that is not discernible on the scales of Fig.~\ref{fig:ledger}. The characteristic accreting mass entering Eq.~(\ref{eq:dSdtBH}) is modeled as $M_{\rm char}(z)=\max[10^{6}\,\Msun,\ \rho_{\rm BH}(<z)/n_{\rm eff}]$, with $n_{\rm eff}=10^{-3}\,{\rm Mpc}^{-3}$ representing the effective comoving density of the black holes that dominate accretion; the band $n_{\rm eff}\in[3\times10^{-4},3\times10^{-3}]$ is propagated. Finally, because JWST-era census studies suggest that the accretion density at $z\gtrsim5$ may exceed pre-JWST extrapolations by one to two orders of magnitude if the little red dots contain accreting black holes \cite{LRDreview2026,Greene2024}, we include an ``LRD bracket'' in which $\psi_{\rm BH}$ is enhanced by a factor of $30$ (band $10$--$100$) at $z\gtrsim5$.

Three validation checks anchor the pipeline. The inferred comoving CMB entropy density is $4.4\times10^{76}\,\kB\,{\rm Mpc}^{-3}$, compared with $4.5\times10^{76}$ in Ref.~\cite{EganLineweaver2010}. The present-day SMBH horizon entropy density is $1.9\times10^{91}\,\kB\,{\rm Mpc}^{-3}$, compared with $2.8\times10^{91}$ in the same census. The cosmic event horizon entropy is $2.95\times10^{122}\,\kB$, compared with $(2.99\pm0.03)\times10^{122}$ in Ref.~\cite{Profumo2024}.

\subsection{Results}
\label{sec:ledgerresults}

\begin{figure}[t]
	\centering
	\includegraphics[width=0.78\textwidth]{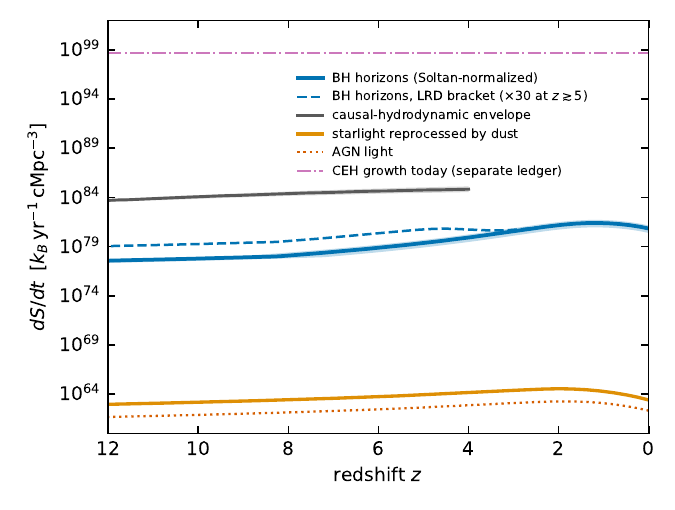}
	\caption{The entropy ledger of the Universe: comoving entropy production rate by channel. The black hole horizon channel (blue; band spans the Soltan normalization, dashed line the LRD bracket) exceeds the radiative channels (orange) by $16$--$17$ orders of magnitude at every epoch after the first seeds. The gray band is the causal-hydrodynamic envelope of Sec.~\ref{sec:envelope}, defined for $4\le z\le12$. The dash-dotted line marks the present growth rate of the cosmic event horizon, kept in a separate ledger (Sec.~\ref{sec:discussion}).}
	\label{fig:ledger}
\end{figure}

\begin{table}[t]
	\centering
	\small
	\begin{tabular}{r c c c c c c}
		\toprule
		$z$ & $\psi_{\rm BH}$ & $M_{\rm char}$ & $\dot S_{\rm BH}$ & $\dot S_{\rm BH}^{\rm LRD}$ & $\dot S_\gamma$ & $F$ (fid.\,--\,LRD) \\
		& $[\Msun\,{\rm yr^{-1}\,Mpc^{-3}}]$ & $[\Msun]$ & \multicolumn{3}{c}{$[\kB\,{\rm yr^{-1}\,Mpc^{-3}}]$} & \\
		\midrule
		12 & $1.8\times10^{-6}$ & $10^{6}$ & $3.8\times10^{77}$ & $1.1\times10^{79}$ & $9.5\times10^{62}$ & $7.5\times10^{-7}$\,--\,$2.3\times10^{-5}$ \\
		10 & $2.9\times10^{-6}$ & $10^{6}$ & $6.1\times10^{77}$ & $1.8\times10^{79}$ & $1.5\times10^{63}$ & $5.2\times10^{-7}$\,--\,$1.6\times10^{-5}$ \\
		8  & $5.2\times10^{-6}$ & $1.1\times10^{6}$ & $1.2\times10^{78}$ & $3.7\times10^{79}$ & $2.8\times10^{63}$ & $5.0\times10^{-7}$\,--\,$1.5\times10^{-5}$ \\
		6  & $1.1\times10^{-5}$ & $3.4\times10^{6}$ & $7.6\times10^{78}$ & $2.3\times10^{80}$ & $5.7\times10^{63}$ & $1.7\times10^{-6}$\,--\,$5.0\times10^{-5}$ \\
		4  & $2.7\times10^{-5}$ & $1.5\times10^{7}$ & $8.4\times10^{79}$ & $5.5\times10^{80}$ & $1.5\times10^{64}$ & $1.2\times10^{-5}$\,--\,$8.1\times10^{-5}$ \\
		2  & $6.9\times10^{-5}$ & $1.1\times10^{8}$ & $1.5\times10^{81}$ & --- & $3.7\times10^{64}$ & --- \\
		1  & $4.5\times10^{-5}$ & $2.6\times10^{8}$ & $2.5\times10^{81}$ & --- & $2.0\times10^{64}$ & --- \\
		0  & $7.8\times10^{-6}$ & $4.2\times10^{8}$ & $6.9\times10^{80}$ & --- & $2.6\times10^{63}$ & --- \\
		\bottomrule
	\end{tabular}
	\caption{The entropy ledger and the efficiency $F(z)$ (fiducial to LRD-bracket range; the envelope, and hence $F$, is defined for $z\ge4$ where atomic-cooling halos are the relevant greedy sites). Full parameter bands in Appendix~\ref{app:numerics}.}
	\label{tab:ledger}
\end{table}

Figure~\ref{fig:ledger} and Table~\ref{tab:ledger} present the first result. Once the first seeds have formed, the horizon channel exceeds all radiative channels by a factor $\sim10^{16\text{--}17}$, confirming and extending across cosmic time what the $z=0$ budgets already imply \cite{EganLineweaver2010}: the entropy history of the Universe is dominated by black hole growth, with all other channels contributing only a minor correction.\footnote{A footnote which, for the record, includes all starlight ever emitted, all neutrinos from all supernovae, us, and the reader.} The rate also has a distinct temporal structure: it rises steeply as the accreting population becomes more massive, peaks at $z\approx1.2$ in the fiducial model (with the exact location shifting between $z\approx1$ and $2$ across the $M_{\rm char}$ band), and subsequently declines by a factor of a few toward the present. The peak entropy production rate therefore coincides with the quasar era; the Universe's period of maximal entropy production lies in the past.

The late position of the peak is robust. Its location is exactly independent of $n_{\rm eff}$, which factors out of the shape of $M_{\rm char}\,\psi_{\rm BH}$, while a ``downsizing'' variant in which the accreting mass saturates at $(1\text{--}5)\times10^{8}\,\Msun$ shifts it only within $z\simeq1.2$--$1.9$, bounded from above by the peak of the accretion history itself at $z\simeq1.9$. None of the variants considered places the peak near cosmic dawn. Section~\ref{sec:physics} explains why this result is structurally expected.

\section{The causal-hydrodynamic envelope}
\label{sec:envelope}

\subsection{Definition}
\label{sec:envdef}

Given Fig.~\ref{fig:ledger}, the natural question is not whether black hole growth dominates entropy production (it does, by sixteen orders of magnitude), but how closely the realized growth approaches the fastest growth permitted by physics; in other words, how efficiently does the Universe produce entropy? We define the \emph{causal-hydrodynamic envelope} as the entropy production of a population in which every atomic-cooling halo channels its gas into a single black hole at the maximum sustained rate allowed by gravity and gas dynamics,
\begin{equation}
	\Mdot_{\rm env} \;\simeq\; \frac{c_s^{3}}{G} \;=\; 0.1\text{--}0.4\;\Msun\,{\rm yr}^{-1}
	\qquad (T\simeq6000\text{--}12000\ {\rm K}),
	\label{eq:envelope}
\end{equation}
with no radiative throttling: the flow is assumed to be obscured and radiatively inefficient, so that the Eddington limit does not constrain the growth. Equation~(\ref{eq:envelope}) is the standard feeding rate for monolithic collapse in halos above the atomic-cooling threshold~\cite{BegelmanVolonteriRees2006,Inayoshi2020}; what is non-standard here is only its use as the growth branch of an extremal trajectory within the entropy ledger. Two caps are imposed: a halo cannot supply more than its baryonic content $f_b M_h$, and the population is drawn from the Sheth--Tormen mass function above the atomic-cooling mass $M_{\rm min}(z)\approx4.6\times10^{7}[(1+z)/11]^{-3/2}\,\Msun$ \cite{BarkanaLoeb2001,ShethTormen1999,Diemer2018}.

The distinction must be stated clearly: the envelope is determined by dynamics, not by thermodynamics. Gravity and gas physics define it; the ledger only identifies it as extremal and measures the distance between it and the realized history. This division of roles is central to the interpretation of the paper. Thermodynamics supplies bounds, envelopes, ledgers, and efficiencies, whereas the rates are set by dynamics; no entropic principle is introduced into Eq.~(\ref{eq:envelope}).

\subsection{Envelope masses versus Eddington tracks}
\label{sec:envmasses}

\begin{figure}[t]
	\centering
	\includegraphics[width=0.78\textwidth]{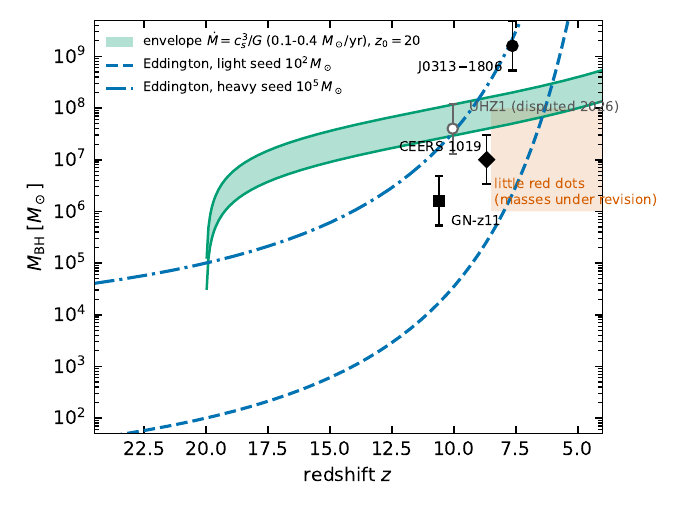}
	\caption{The causal-hydrodynamic envelope (green band: $\Mdot=0.1$--$0.4\,\Msun\,{\rm yr^{-1}}$ from $z_0=20$) against Eddington-limited tracks ($t_{\rm Sal}=50$ Myr) from light ($10^2\,\Msun$) and heavy ($10^5\,\Msun$) seeds, and against the 2026 observational landscape: the $z=7.64$ quasar J0313$-$1806 \cite{Wang2021}, the low-mass AGN GN-z11 \cite{Maiolino2024} and CEERS 1019, the disputed UHZ1 \cite{UHZ1reanalysis2026,Natarajan2024}, and the little red dot mass range, explicitly marked as under revision \cite{LRDreview2026,Overmassive2026}.}
	\label{fig:envelope}
\end{figure}

Figure~\ref{fig:envelope} compares the envelope with Eddington-limited growth. Between $z=20$ and $z=7.64$, the Universe provides $499$ Myr, almost exactly ten Salpeter $e$-folds: a light seed of $10^{2}\,\Msun$ reaches only $2\times10^{6}\,\Msun$, approximately three orders of magnitude below J0313$-$1806, which is the standard early-growth tension~\cite{Inayoshi2020,Volonteri2021}. By contrast, the envelope reaches $M_{\rm BH}\sim(1\text{--}8)\times10^{7}\,\Msun$ by $z\approx8$--$10$ without being limited by the Eddington rate, and subsequently gives way to Eddington-limited growth in more massive halos during the final rise to $10^{9}\,\Msun$. The extremal trajectory is not astrophysically exotic; it corresponds closely to the heavy-seed, direct-collapse scenario developed in the literature for independent reasons~\cite{BegelmanVolonteriRees2006,Inayoshi2020}.

A structural inversion lies at the conceptual center of the paper. The envelope requires the gas \emph{not} to cool efficiently: molecular cooling must be suppressed, otherwise the cloud fragments into stars and monolithic collapse is prevented \cite{BegelmanVolonteriRees2006}. Locally, fragmentation into stars is the more dissipative and entropy-producing path, whereas monolithic collapse is comparatively quiet. Globally, however, the corresponding entropy yields differ by sixteen orders of magnitude in the opposite direction. Reaching the global maximum therefore requires foregoing the local one. Dynamics acting through local physical processes has no mechanism for evaluating that global trade-off: the local channel is selected in nearly every case, and the Universe consequently fills with stars.\footnote{This is the sense in which structure formation is a greedy algorithm. The one systematic exception, halos bathed in just enough Lyman--Werner radiation to keep H$_2$ dissociated, achieves the global strategy precisely by having its local dissipation channel externally disabled.}

\subsection{The envelope has the face of a little red dot}
\label{sec:LRD}

An object evolving along the envelope grows within an optically thick gaseous cocoon, radiates inefficiently, produces few hard photons, and presents a cool, red photosphere at a temperature of a few thousand kelvin. This corresponds closely, point by point, to the observed phenomenology of JWST little red dots: compact sources with V-shaped spectral energy distributions, broad Balmer lines, X-ray and radio weakness even in stacked observations, and abundances well above pre-JWST AGN extrapolations~\cite{LRDreview2026,Greene2024}. In addition, ``black hole star'' models propose precisely an accreting black hole embedded within a dense gaseous envelope~\cite{BHstar2025}. The envelope masses, $10^{7\pm1}\,\Msun$ at $z\approx8$--$10$ with $M_{\rm BH}/M_\star$ of order $0.1$--$1$ (the black hole assembles before most of the host's stellar component), lie within the reported LRD range.

We present this relation as consistency rather than derivation for two reasons. Conceptually, the envelope is defined by extremality, and the resemblance between the extremal trajectory and the observed population does not establish that extremality explains why the population exists; the operative causes remain astrophysical selection effects, including Lyman--Werner backgrounds, streaming velocities, and merger histories, which act only in a small fraction of halos. Observationally, the relevant masses remain under active revision: the direct-collapse interpretation of UHZ1 was challenged in 2026~\cite{UHZ1reanalysis2026}, and 2026 reanalyses argue that the purportedly ``overmassive'' black holes may be substantially less massive than initially reported~\cite{Overmassive2026}. Conversely, a direct dynamical mass measurement in one little red dot, based on spatially resolved gas kinematics rather than on line widths, finds a genuinely overmassive black hole~\cite{DirectMass2025}; the ongoing revision therefore cuts in both directions. The consistency claim would survive a downward revision of the LRD masses by an order of magnitude. What would instead undermine the envelope interpretation is evidence that early growth is radiatively \emph{efficient} and constrained by the Eddington limit, because in that case the envelope would not describe any realized population. This constitutes the falsifiable aspect of the present section.

\section{The efficiency of cosmic entropy production}
\label{sec:efficiency}

\begin{figure}[t]
	\centering
	\includegraphics[width=0.72\textwidth]{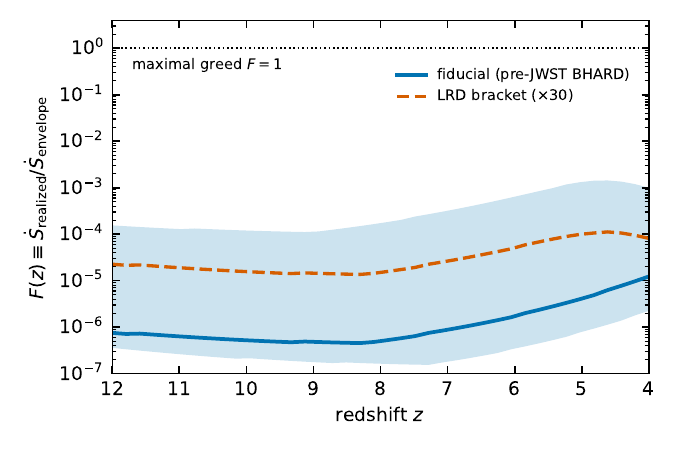}
	\caption{The efficiency $F(z)$: realized entropy production over the causal-hydrodynamic envelope, for the fiducial (Soltan-normalized) and LRD-bracket accretion histories. The shaded band propagates the normalization, $n_{\rm eff}$, and envelope-rate uncertainties. The Universe runs four to seven orders of magnitude below maximal greed at all epochs.}
	\label{fig:efficiency}
\end{figure}

The ratio of the quantities defined in the two preceding sections yields the central dimensionless measure of this paper,
\begin{equation}
	F(z) \;\equiv\; \frac{(dS/dt)_{\rm realized}}{(dS/dt)_{\rm envelope}},
	\label{eq:F}
\end{equation}
shown in Fig.~\ref{fig:efficiency} for $4\le z\le12$. In the fiducial history, $F$ ranges from $5\times10^{-7}$ to $10^{-5}$; the LRD bracket raises this range to $1.5\times10^{-5}$ to $10^{-4}$. Within the systematic band, $F$ is consistent with an approximately constant value of $10^{-5\pm1.5}$ across the full interval: the Universe produces entropy at a persistently small fraction of the available envelope. Two companion quantities extend this comparison to the present epoch, where the envelope would require reformulation for cluster environments: the mass efficiency $F_M=\rho_{\rm BH,0}/\rho_b\approx7\times10^{-5}$, and the entropy efficiency relative to the complete collapse of the baryons, $F_S\sim5\times10^{-6}$ to $3\times10^{-4}$ depending on the assumed remnant masses.

If the JWST-era censuses are confirmed, their implication in this framework is specific: the early Universe operated approximately thirty times closer to its envelope than pre-JWST extrapolations indicated, making cosmic dawn the epoch at which realized entropy production came closest to the causally available envelope. Even then, $F\lesssim10^{-4}$. The gap is neither closed nor closely approached.

\section{Three no-go results}
\label{sec:nogo}

The purpose of the ledger and the envelope is to render extremal claims testable. Three such claims fail immediately.

\subsection{Boltzmann weighting of channels}
\label{sec:boltzmann}

Suppose that channel selection were governed by any weight monotonic in the entropy gain, with $P\propto e^{\Delta S/\kB}$ as the most direct, if inappropriate, example. For $10^{5}\,\Msun$ of gas, the black hole channel yields $\Delta S = 1.05\times10^{87}\,\kB$; the stellar channel, after integrating the radiative output of a normal stellar population over $10$ Gyr and degrading all of it to a dust temperature of $30$ K, which is already a generous accounting, yields $3.0\times10^{70}\,\kB$. Any Boltzmann-type weighting would therefore select collapse with probability $1-e^{-10^{87}}$. This is incompatible with the observation that gas fragments into stars in essentially every halo, so the proposal fails at leading order, without requiring further observational detail. The factor left unrealized, $3.5\times10^{16}$ per episode, is perhaps the single most instructive number in the paper.

\subsection{The dissipative-adaptation transplant}
\label{sec:england}

A more sophisticated proposal invokes England's dissipative adaptation \cite{England2013,England2015}: driven systems statistically favor histories that dissipate more work. The theorem, however, concerns transition probabilities between mesostates connected by thermal fluctuations, under conditions that include a heat bath, local detailed balance, and time-scale separation \cite{Seifert2012}. None of these conditions relates the ``stellar'' and ``black hole'' macrostates of a protogalactic cloud: they are not connected by fluctuations on any relevant timescale, there is no heat bath in the required sense, and self-gravity invalidates the extensivity assumed by the formalism. No gravitational analogue of the theorem is presently available, and this paper does not provide one; the transplantation therefore fails categorically rather than numerically.\footnote{Nothing in this argument diminishes the theorem within its proper domain. The point is that a result formulated for driven mesoscopic systems cannot be transferred directly to cosmological structure formation.}

\subsection{Strong maximum entropy production}
\label{sec:mepp}

The maximum-entropy-production principle, in its strong cosmological interpretation that structure forms so as to maximize the entropy production rate, inherits both failures and introduces an additional one. Near equilibrium, the only established theorem selects \emph{minimum} entropy production~\cite{Prigogine1947}; far from equilibrium, no general variational principle exists, proposed derivations have not withstood scrutiny~\cite{Martyushev2021,GrinsteinLinsker2007}, and the cosmological record is inconsistent with maximization: $F\sim10^{-5}$, stars form in nearly every eligible system, and the envelope is realized only where local dissipation is externally suppressed (Sec.~\ref{sec:envmasses}). The greedy inversion is the structural obstruction: strong MEPP requires the dynamics to forgo local optima in favor of a global one, whereas local field theories do not perform such a comparison. A constrained selection statement may still survive---for example, that among the steady states accessible to a given halo, the realized state is the most dissipative---but its constraints must be specified ex ante, and its cosmological content, if any, remains to be established. We know of no existing formulation that is compatible with Fig.~\ref{fig:efficiency}.

\section{Why the ledger looks the way it does}
\label{sec:physics}

The results above are quantitative, and two of them run counter to a reasonable thermodynamic intuition. This section argues that they are not merely empirical outcomes but consequences of the structure of the problem. We examine them in some detail because the basis of their plausibility is itself physical: each result decomposes into one dynamical fact and one property of horizon entropy, and these decompositions are exact.


The least intuitive result is the location of the peak. A Universe that begins as far from its entropic ceiling as causality permits, and whose disequilibrium in the growing-gap sense only widens~\cite{Frautschi1982,Layzer1975}, might be expected to produce entropy most rapidly at early times, when the thermodynamic drive is greatest. The ledger contradicts this expectation, and the reason is arithmetic before it is physical.

Horizon entropy production is not proportional to the accretion flow. It is the flow weighted by the mass of the accretor, Eq.~(\ref{eq:dSdtBH}): $\dot S\propto M\Mdot$. The resulting structure closely resembles compound interest. The ledger grows as capital times income, while the capital is itself accumulated income.\footnote{Equivalently, $\dot S\propto d(M^2)/dt$: the ledger tracks the second moment of the black hole mass function, and second moments live in the high-mass tail. In a hierarchical cosmology the tail is built last.} Cosmic income, $\psi_{\rm BH}(z)$, peaks at cosmic noon for standard dynamical reasons: the massive, gas-rich halos that feed quasars are not assembled until $z\sim2$, after which the supply is suppressed by feedback and by $\Lambda$. Cosmic capital, $M_{\rm char}(z)$, can only increase because black holes retain their accumulated mass. The product must therefore peak at, or after, the income peak, and not before it. An early peak is not merely disfavored by this structure; it is excluded by it.

It is useful to distinguish three different ``peaks'', because the phrase \emph{black hole formation} can obscure their separation. The formation of the first black holes and heavy seeds peaks early, at $z\sim10$--$25$. The accretion of mass peaks at cosmic noon, $z\approx2$. Entropy production peaks last, because a solar mass produces ten thousand times more entropy when accreted by a $10^{9}\,\Msun$ black hole at $z=1$ than by a $10^{5}\,\Msun$ black hole at $z=10$. The Universe therefore produces most of its entropy neither when its black holes are born nor primarily when they accrete most rapidly, but when the most massive accretors have finally assembled.

The failure of the early-peak intuition is itself informative. That intuition follows a linear-response picture: flux is proportional to force, so maximal disequilibrium should generate maximal production. Linear response, however, presumes a fixed conductance, whereas here the effective conductance is the assembled black hole population, which begins near zero and is created by the same flux that it subsequently carries. The early Universe is rich in thermodynamic drive but poor in the structures capable of converting it into horizon entropy. Production is small at high $z$ not because the drive is weak, but because the relevant accretors have not yet formed, and their construction is the slow process of hierarchical assembly, further limited by the Eddington rate and the Salpeter time. Entropy production in this Universe is autocatalytic, and autocatalytic growth is correspondingly weighted toward later times.

The intuition is therefore not entirely wrong, but applies to a different history. It is correct for the envelope of Sec.~\ref{sec:envelope}. A population able to convert baryons at the causal-hydrodynamic rate front-loads its growth: it consumes the available gas in each halo within a few hundred Myr of formation and thereafter grows mainly through mergers. An early peak characterizes maximal greed, whereas a late peak is the outcome of a greedy algorithm. The displacement of the peak from cosmic dawn to cosmic noon is $F(z)$ expressed temporally rather than as a rate ratio, and the expectation of an early maximum is precisely the null hypothesis rejected by the ledger.


The dominance of the horizon channel may appear to reflect the amount of energy processed, but it is principally a consequence of temperature. At cosmic noon ($z \sim 2$, the joint peak of cosmic star formation and quasar activity), the energy processed by black hole accretion and that radiated by stars are, through a mild numerical coincidence of the Soltan normalization, comparable: $6.9\times10^{-5}$ against $9.2\times10^{-5}\,\Msun c^{2}\,{\rm yr^{-1}\,Mpc^{-3}}$, a ratio of $0.74$. The difference lies in the entropy produced per unit energy. Entropy per unit energy is $1/T$; dust re-radiates starlight at $35$ K, whereas a $10^{8}\,\Msun$ horizon absorbs at $T_{\rm H}=6\times10^{-16}$ K, so a joule produces $T_{\rm dust}/T_{\rm H}=5.7\times10^{16}$ times more entropy in the horizon channel. The product of the two ratios, $4.2\times10^{16}$, reproduces the measured separation between the two curves in Fig.~\ref{fig:ledger} at $z=2$, namely $4.1\times10^{16}$: the sixteen orders of magnitude are accounted for by the temperature ratio alone.\footnote{The identity behind this is $dS_{\rm hor}/dE=1/T_{\rm H}$, the first law applied to the horizon; Eq.~(\ref{eq:dSdtBH}) is that identity in disguise.} Black holes dominate the entropy budget not because they process more energy than starlight, since they process slightly less, but because they are the coldest objects in the Universe, and for horizons temperature and size are inversely related, $T_{\rm H}\propto1/M$. In this sense, the ledger is a thermometer read in reverse. The late peak discussed above reappears from this perspective: the accumulation of capital is also a cooling process, and entropy production is greatest when the coldest reservoirs are both sufficiently massive and still being supplied.


These considerations allow $F(z)$ to be interpreted as encoding two facts. Its smallness measures the rarity of the anti-dissipative path: the envelope is realized only where monolithic collapse survives, namely where local cooling and fragmentation have been externally suppressed. Pre-JWST models placed the required selection at roughly one halo in $10^{6}$--$10^{8}$ \cite{Inayoshi2020}, whereas the inferred little red dot abundances, if confirmed, would relax this to approximately one in $10^{3}$--$10^{4}$. Gas that does not follow the envelope is instead distributed into stars, whose entropic yield is smaller by the factor discussed above; the smallness of $F$ is therefore the score obtained by a greedy algorithm when judged against a globally optimizing one. Its relative flatness indicates that the numerator and denominator evolve on similar clocks: both realized accretion and the available envelope track the assembly of atomic-cooling halos, so their ratio changes by less than an order of magnitude while the rates themselves increase by two to three. $F$ is therefore an effective greediness coefficient for the Universe, and its inferred behavior, $10^{-5\pm1.5}$ with no epoch approaching unity, states the paper's central negative result in positive form: the quantity is well defined, small, and comparatively stable. In this language, the principal thermodynamic implication of the JWST era is specific. Cosmic dawn may have operated a factor $\sim30$ closer to the envelope than pre-JWST censuses implied, while remaining four orders of magnitude below it.


The meaning of these numbers is that they enforce a division of labor that entropic teleologies tend to obscure. The second law fixes the direction of every process in Fig.~\ref{fig:ledger} and forbids none of them; the rates are determined by collapse times, cooling channels, angular momentum transport, and radiative throttling, none of which is set by the entropy ledger. In the analogy of an institution, the second law audits each transaction and verifies the accounts but does not determine strategy. This is its role in cosmic history, and the result of Sec.~\ref{sec:nogo} is that assigning it a strategic role fails by sixteen orders of magnitude. The declining tail of the ledger is then unsurprising: after cosmic noon, the Universe contains immense accumulated black hole mass but a declining accretion supply, and its period of maximal entropy production has passed.\footnote{The analogy is imperfect in one respect: this rentier's capital is also its furnace, and in some $10^{100}$ years it will have evaporated. No metaphor survives contact with Hawking radiation.} The opposite result would have required explanation. An entropy-production peak at cosmic dawn, an $F$ of order unity, or gas systematically foregoing the stellar channel in favor of monolithic collapse would constitute genuine evidence for entropic selection. Their absence is informative precisely because these are the signatures that such a principle would be expected to produce.

\section{Discussion}
\label{sec:discussion}


The analysis above excludes the cosmic event horizon from the accounting, and that exclusion is consequential. If the Gibbons--Hawking entropy of the CEH \cite{GibbonsHawking1977} is admitted as a bona fide entry, it exceeds the total black hole contribution by eighteen orders of magnitude ($3\times10^{122}$ against $3\times10^{104}\,\kB$ \cite{EganLineweaver2010,Profumo2024}), while its present growth rate, $\sim5\times10^{111}\,\kB\,{\rm yr}^{-1}$, exceeds the entire black hole channel by a comparable margin. In that ledger, the Universe's most effective entropy-producing process is simply expansion, black holes become a negligible correction, and the teleological question reduces to the asymptotic approach to de Sitter space. The CEH is, however, observer-dependent,\footnote{An uncomfortable property for the terminal entry of a ledger.} and one may reasonably decline to count it, as Penrose would~\cite{Penrose1989}. We take no position on that choice; the point is structural. \emph{The teleology is not robust to the choice of ledger}: with the CEH, the Universe maximizes through expansion; without it, through collapse. A purported ``goal'' that reverses under a bookkeeping convention cannot constitute a robust physical principle. This observation is at least as damaging to entropic teleology as the numerical no-go results, and requires no additional modeling.


More fundamentally, the smooth initial state, the Past Hypothesis in Albert's sense \cite{Albert2000}, with Penrose's vanishing initial Weyl curvature as its geometric expression \cite{Penrose1979}, states that the Universe began without realizing the largest entropy gain then permitted by causality: horizon-mass black holes throughout the radiation era, whose entropy by $t\sim1$ s would have exceeded everything accumulated since. Whatever explains this initial condition, and we offer no explanation here, its role in the present ledger is clear: it constitutes the primordial prohibition on immediate gravitational saturation and thereby makes possible a prolonged history of structure formation \cite{Frautschi1982,PatelLineweaver2017,Banks2021}. Recent thermodynamic work addresses the homogeneity and flatness problems directly \cite{TurokBoyle2024}; none of the present accounting depends on how that debate is resolved. Low initial entropy and the small efficiency quantified by $F(z)$ are therefore not wholly separate puzzles: the Universe begins far below the causally available envelope and subsequently approaches it at only $10^{-5}$ of the allowed rate.

\subsection{Relation to earlier proposals}
\label{sec:related}

Frautschi's classic analysis \cite{Frautschi1982} established the growing entropy gap and identified black hole formation as the dominant channel; the present paper adds a reconstructed production history, a causal-hydrodynamic envelope, and an associated efficiency. Lineweaver's proposal that MEPP might be tested through accretion efficiency \cite{Lineweaver2014} receives a negative answer here at the population level, irrespective of what may hold for individual disks. The causal entropic principle of Bousso et al.~\cite{Bousso2007} weights vacua by entropy production within causal diamonds while deliberately excluding horizon entropy; the present ledger shows how consequential that exclusion is, because including horizons increases the relevant weighting by $\sim10^{16}$ and transfers its dominance to black holes. Smolin's cosmological natural selection \cite{Smolin1992} reaches a superficially similar conclusion, namely that universes producing many black holes are favored, but does so through reproduction and selection rather than thermodynamic preference; the present results are neutral with respect to that proposal, although the smallness of $F$ suggests that any selection for black hole production would operate under strong constraints or with limited efficiency. Finally, the holographic cosmology of Banks and Fischler \cite{Banks2021,BanksFischler2024}, in which the early Universe is represented as a maximal-entropy black hole gas, may be viewed as the counterfactual in which the primordial prohibition was not imposed; the observed Universe records the opposite history. The only route by which entropic language might acquire genuine dynamical force, namely the interpretation of gravity itself as thermodynamics \cite{Jacobson1995,ChircoLiberati2010,Bianconi2025,Bianconi2026,DorauMuch2026}, is not addressed by the no-go results above. Two remarks locate the present paper relative to that program, taking gravity-from-entropy theory as its currently most developed representative. First, the ledger constructed here is unchanged under that framework by construction: the theory reduces to general relativity with an emergent dark-energy term in its low-energy, small-curvature limit \cite{Bianconi2025}, recovers the Schwarzschild area law \cite{BianconiEntropy2025}, and reproduces the $H^{-2}$ de Sitter scaling \cite{Bianconi2026}, while every quantity in Secs.~\ref{sec:ledger}--\ref{sec:efficiency} is evaluated in that limit. Second, such a theory aims to provide precisely what Sec.~\ref{sec:ledger} lacks: a local, volumetric gravitational entropy defined independently of horizons. Its thermodynamic formulation has so far been developed only for homogeneous backgrounds, where the local entropy density decreases while the total entropy remains nondecreasing \cite{Bianconi2026}, a background-level analogue of the local--global tension discussed in Sec.~\ref{sec:mepp}. Whether the clustering contribution to such an entropy increases during structure formation is, to our knowledge, an open and well-posed question, and constitutes a natural point of contact between that program and the present ledger. Computing it for a perturbed Friedmann metric, and comparing the result with the Weyl-based proposal of Ref.~\cite{CET2013} and with the gravitational dissipation of Chirco and Liberati \cite{ChircoLiberati2010}, whose shear-squared form connects the same set of ideas to the Penrose boundary condition, could determine which candidate, if any, satisfies a second law during structure formation. We leave these questions open.

\subsection{What would change these conclusions}
\label{sec:falsifiability}

The ledger in Sec.~\ref{sec:ledger} is robust at the order-of-magnitude level under every input uncertainty considered; overturning the $10^{16}$ dominance of the horizon channel would require a failure of the Bekenstein--Hawking entropy assignment itself. The no-go results of Sec.~\ref{sec:nogo} are comparably insensitive. The consistency claim concerning the envelope in Sec.~\ref{sec:LRD} is the most exposed component: it would be strengthened if LRD spectroscopy confirmed abundant, obscured, radiatively inefficient super-Eddington growth, and weakened if the population were instead resolved into compact starbursts or if early accretion proved to be efficient and Eddington-throttled. $F(z)$ would then shift by one to two orders of magnitude, which does not alter the conclusion when set against a seven-order gap. The central results therefore rest on robust order-of-magnitude separations rather than on the presently unsettled high-$z$ mass estimates.

\section{Conclusions}
\label{sec:conclusions}

Four points summarize this paper: (i) The Universe produces entropy predominantly in black hole horizons, at $10^{16\text{--}17}$ times the rate of all radiative channels combined. (ii) This production peaked at $z\approx1$--$2$, rather than at cosmic dawn, and is now declining. (iii) The fastest trajectory permitted by causality and hydrodynamics was available from the first atomic-cooling halos, reaches $10^{7\pm1}\,\Msun$ by $z\approx9$, and closely resembles the inferred properties of a little red dot. And (iv) the Universe follows that trajectory with an efficiency $F\sim10^{-5\pm1.5}$ throughout the redshift interval and parameter brackets considered.

Each result has a transparent physical interpretation (Sec.~\ref{sec:physics}), and none is obscure once its underlying structure is identified. The dominance is primarily a temperature ratio rather than a difference in energy processing: the energy fluxes through horizons and through starlight are comparable, while $1/T$ accounts for the large entropy contrast. The late peak follows a compound-interest structure: horizon entropy production weights income by capital, $\dot S\propto M\Mdot$, and the capital is assembled late; a maximally greedy universe would instead front-load its production, as the envelope does but the realized history does not. The smallness of $F$ reflects the rarity of the anti-dissipative route: the largest global entropy yield requires the suppression of local dissipation, whereas local dynamics selects the locally accessible channel in nearly every case. Entropy production therefore provides an accurate account of what the Universe does, but not a sufficient explanation of why it does so. The second law constrains and records every process in cosmic history; it does not determine the dynamical strategy.

The resulting program is concrete. The ledger can be refined by replacing the Soltan-normalized shape with full bolometric luminosity functions and active black hole mass functions; the envelope can be computed halo by halo; and $F(z)$ at cosmic dawn is, in effect, now being constrained by spectroscopic studies of little red dots, the first population that may be following the envelope while directly observed. If their inferred masses and abundances survive the current revision, the early Universe operated thirty times closer to maximal greed than pre-JWST estimates implied, while remaining four orders of magnitude below it. Structure formation behaves as a greedy algorithm that systematically forgoes a global maximum inaccessible to local dynamics. A defensible thermodynamic cosmology therefore consists in accounting for what remains unrealized: sixteen orders of magnitude per fragmenting cloud, four to seven orders in rate, and five times the present total by the first second of cosmic time.

\subsection*{Acknowledgments}
F.~Izaurieta acknowledges support from ANID-FONDECYT Regular grant 1262414, C.~Quinzacara from ANID-FONDECYT grant 11231238, and O.~Valdivia from ANID-FONDECYT grant 1251523, all from the Government of Chile. F.~Izaurieta is also grateful for the warm hospitality at ICEN Iquique: the dark night skies of the Atacama Desert, together with the entropic wild seas of northern Chile, were the starting point of this work.

\subsection*{AI disclosure statement}
This article was created and composed by humans. However, Grammarly AI was utilized to enhance many grammatical, spelling, and stylistic errors (paragraphs organization) commonly made by non-native English authors. After using this tool, the authors reviewed and edited the content as needed and take full responsibility for the content of the published article.

\appendix

\section{Numerical details}
\label{app:numerics}

Cosmological parameters are taken from Planck 2018 ($h=0.6736$, $\Omega_m=0.3153$, $\Omega_b=0.0493$). The halo mass function is the Sheth--Tormen form \cite{ShethTormen1999}, evaluated with \textsc{colossus} \cite{Diemer2018} using the FOF definition and integrated from the atomic-cooling mass $M_{\rm min}(z)$ (Barkana--Loeb with $T_{\rm vir}=10^4$ K, $\mu=0.6$ \cite{BarkanaLoeb2001}) to $10^{13.5}\,\Msun$. For each halo, the envelope mass is $M_{\rm BH}=\min[\Mdot_{\rm env}\,\Delta t,\ f_b M_h]$ with $\Delta t=t(z)/2$ and $\Mdot_{\rm env}$ set to zero once the baryonic budget is exhausted. The parameter bands propagated in Figs.~\ref{fig:ledger} and \ref{fig:efficiency} are $\rho_{\rm BH,0}\in[2.5,6.0]\times10^{5}\,\Msun\,{\rm Mpc^{-3}}$; $n_{\rm eff}\in[3\times10^{-4},3\times10^{-3}]\,{\rm Mpc^{-3}}$; $\Mdot_{\rm env}\in[0.1,0.4]\,\Msun\,{\rm yr^{-1}}$; LRD boost $\in[10,100]$ at $z\gtrsim5$. 

\section{The cosmic event horizon ledger}
\label{app:CEH}

The proper CEH radius $R_{\rm CEH}(z)=a(z)\int_{t(z)}^{\infty} c\,dt'/a(t')$ evaluates today to $5.08$ Gpc, yielding $S_{\rm CEH}=\pi R_{\rm CEH}^2 c^3 \kB/G\hbar = 2.95\times10^{122}\,\kB$ and a present growth rate of $5.3\times10^{111}\,\kB\,{\rm yr}^{-1}$. Dividing this rate by the comoving volume of the observable Universe ($1.12\times10^{13}\,{\rm Mpc}^{3}$) gives $4.7\times10^{98}\,\kB\,{\rm yr^{-1}\,Mpc^{-3}}$, although the legitimacy of this operation is precisely what Sec.~\ref{sec:discussion} questions. As the Universe approaches de Sitter, $S_{\rm CEH}$ saturates at $3\pi c^5 \kB/(G\hbar\Lambda)\approx 3.1\times10^{122}\,\kB$ and the growth rate tends to zero; the CEH ledger, like the black hole ledger, has therefore passed its peak.


\begin{thebibliography}{99}
\small

\bibitem{EganLineweaver2010} C.~A. Egan and C.~H. Lineweaver, \emph{A larger estimate of the entropy of the universe}, Astrophys. J. \textbf{710}, 1825 (2010) [arXiv:0909.3983].

\bibitem{Profumo2024} S. Profumo et al., \emph{A new census of the Universe's entropy}, J. Cosmol. Astropart. Phys. \textbf{09} (2025) 049 [arXiv:2412.11282].

\bibitem{Frautschi1982} S. Frautschi, \emph{Entropy in an expanding universe}, Science \textbf{217}, 593 (1982).

\bibitem{LineweaverEgan2008} C.~H. Lineweaver and C.~A. Egan, \emph{Life, gravity and the second law of thermodynamics}, Phys. Life Rev. \textbf{5}, 225 (2008).

\bibitem{Lineweaver2014} C.~H. Lineweaver, \emph{The entropy of the universe and the maximum entropy production principle}, in R.~C. Dewar et al. (eds.), \emph{Beyond the Second Law}, Springer (2014), ch. 22.

\bibitem{PatelLineweaver2017} V.~M. Patel and C.~H. Lineweaver, \emph{Solutions to the cosmic initial entropy problem without equilibrium initial conditions}, Entropy \textbf{19}, 411 (2017) [arXiv:1708.03677].

\bibitem{Layzer1975} D. Layzer, \emph{The arrow of time}, Sci. Am. \textbf{233}, No.~6, 56 (1975).

\bibitem{Wang2021} F. Wang et al., \emph{A luminous quasar at redshift 7.642}, Astrophys. J. Lett. \textbf{907}, L1 (2021) [arXiv:2101.03179].

\bibitem{Maiolino2024} R. Maiolino et al., \emph{A small and vigorous black hole in the early Universe}, Nature \textbf{627}, 59 (2024) [arXiv:2305.12492].

\bibitem{LRDreview2026} D.~D. Vaida and R.~J. Farber, \emph{Little red dots: the assembly of early supermassive black holes in the JWST era}, Front. Astron. Space Sci. \textbf{13}, 1779045 (2026).

\bibitem{MartyushevSeleznev2006} L.~M. Martyushev and V.~D. Seleznev, \emph{Maximum entropy production principle in physics, chemistry and biology}, Phys. Rep. \textbf{426}, 1 (2006).

\bibitem{Martyushev2021} L.~M. Martyushev, \emph{Maximum entropy production principle: history and current status}, Phys. Usp. \textbf{64}, 558 (2021). 

\bibitem{Greene2024} J.~E. Greene et al., \emph{UNCOVER spectroscopy confirms the surprising ubiquity of active galactic nuclei in red sources at $z > 5$}, Astrophys. J. \textbf{964}, 39 (2024) [arXiv:2309.05714].

\bibitem{BHstar2025} R.~P. Naidu et al., \emph{A ``black hole star'' reveals the remarkable gas-enshrouded hearts of the little red dots}, arXiv:2503.16596 (2025);
A. de Graaff et al., \emph{A remarkable Ruby: Absorption in dense gas, rather than evolved stars, drives the extreme Balmer break of a Little Red Dot at $z=3.5$}, Astron. Astrophys. \textbf{701}, A168 (2025) [arXiv:2503.16600].

\bibitem{UHZ1reanalysis2026} F. Zou, E. Gallo, Z. Zuo, E. Hodges-Kluck et al., \emph{Revisiting the claim for a direct-collapse black hole in UHZ1 at $z=10.05$}, arXiv:2603.24893 (2026).

\bibitem{Overmassive2026} V. Rusakov, D. Watson, G.~P. Nikopoulos, G. Brammer et al., \emph{Little red dots as young supermassive black holes in dense ionized cocoons}, Nature \textbf{649}, 574 (2026) [arXiv:2503.16595].

\bibitem{DirectMass2025} I. Juod\v{z}balis, C. Marconcini, F. D'Eugenio, R. Maiolino et al., \emph{A direct black hole mass measurement in a Little Red Dot at the Epoch of Reionization}, arXiv:2508.21748 (2025).

\bibitem{MadauDickinson2014} P. Madau and M. Dickinson, \emph{Cosmic star-formation history}, Annu. Rev. Astron. Astrophys. \textbf{52}, 415 (2014).

\bibitem{Soltan1982} A. Soltan, \emph{Masses of quasars}, Mon. Not. R. Astron. Soc. \textbf{200}, 115 (1982).

\bibitem{Shankar2009} F. Shankar, D.~H. Weinberg and J. Miralda-Escud\'e, \emph{Self-consistent models of the AGN and black hole populations: duty cycles, accretion rates, and the mean radiative efficiency}, Astrophys. J. \textbf{690}, 20 (2009) [arXiv:0710.4488].

\bibitem{Shen2020} X. Shen et al., \emph{The bolometric quasar luminosity function at $z=0$--$7$}, Mon. Not. R. Astron. Soc. \textbf{495}, 3252 (2020) [arXiv:2001.02696].

\bibitem{BegelmanVolonteriRees2006} M.~C. Begelman, M. Volonteri and M.~J. Rees, \emph{Formation of supermassive black holes by direct collapse in pre-galactic haloes}, Mon. Not. R. Astron. Soc. \textbf{370}, 289 (2006).

\bibitem{Inayoshi2020} K. Inayoshi, E. Visbal and Z. Haiman, \emph{The assembly of the first massive black holes}, Annu. Rev. Astron. Astrophys. \textbf{58}, 27 (2020) [arXiv:1911.05791].

\bibitem{BarkanaLoeb2001} R. Barkana and A. Loeb, \emph{In the beginning: the first sources of light and the reionization of the universe}, Phys. Rep. \textbf{349}, 125 (2001).

\bibitem{ShethTormen1999} R.~K. Sheth and G. Tormen, \emph{Large-scale bias and the peak background split}, Mon. Not. R. Astron. Soc. \textbf{308}, 119 (1999).

\bibitem{Diemer2018} B. Diemer, \emph{COLOSSUS: a Python toolkit for cosmology, large-scale structure, and dark matter halos}, Astrophys. J. Suppl. \textbf{239}, 35 (2018) [arXiv:1712.04512].

\bibitem{Natarajan2024} P. Natarajan et al., \emph{First detection of an overmassive black hole galaxy UHZ1}, Astrophys. J. Lett. \textbf{960}, L1 (2024) [arXiv:2308.02654].

\bibitem{Volonteri2021} M. Volonteri, M. Habouzit and M. Colpi, \emph{The origins of massive black holes}, Nat. Rev. Phys. \textbf{3}, 732 (2021) [arXiv:2110.10175].

\bibitem{England2013} J.~L. England, \emph{Statistical physics of self-replication}, J. Chem. Phys. \textbf{139}, 121923 (2013).

\bibitem{England2015} J.~L. England, \emph{Dissipative adaptation in driven self-assembly}, Nat. Nanotechnol. \textbf{10}, 919 (2015).

\bibitem{Seifert2012} U. Seifert, \emph{Stochastic thermodynamics, fluctuation theorems and molecular machines}, Rep. Prog. Phys. \textbf{75}, 126001 (2012).

\bibitem{Prigogine1947} I. Prigogine, \emph{\'Etude thermodynamique des ph\'enom\`enes irr\'eversibles}, Desoer, Li\`ege (1947); D. Kondepudi and I. Prigogine, \emph{Modern Thermodynamics}, 2nd ed., Wiley (2014).

\bibitem{GrinsteinLinsker2007} G. Grinstein and R. Linsker, \emph{Comments on a derivation and application of the `maximum entropy production' principle}, J. Phys. A \textbf{40}, 9717 (2007).

\bibitem{GibbonsHawking1977} G.~W. Gibbons and S.~W. Hawking, \emph{Cosmological event horizons, thermodynamics, and particle creation}, Phys. Rev. D \textbf{15}, 2738 (1977).

\bibitem{Penrose1989} R. Penrose, \emph{The Emperor's New Mind}, Oxford University Press (1989).

\bibitem{Albert2000} D.~Z. Albert, \emph{Time and Chance}, Harvard University Press (2000).

\bibitem{Penrose1979} R. Penrose, \emph{Singularities and time-asymmetry}, in S.~W. Hawking and W. Israel (eds.), \emph{General Relativity: An Einstein Centenary Survey}, Cambridge University Press (1979).

\bibitem{Banks2021} T. Banks, \emph{Entropy and black holes in the very early universe}, arXiv:2109.05571 (2021).

\bibitem{TurokBoyle2024} N. Turok and L. Boyle, \emph{Gravitational entropy and the flatness, homogeneity and isotropy puzzles}, Phys. Lett. B \textbf{849}, 138443 (2024); L. Boyle and N. Turok, \emph{Thermodynamic solution of the homogeneity, isotropy and flatness puzzles (and a clue to the cosmological constant)}, Phys. Lett. B \textbf{849}, 138442 (2024).

\bibitem{Bousso2007} R. Bousso, R. Harnik, G.~D. Kribs and G. Perez, \emph{Predicting the cosmological constant from the causal entropic principle}, Phys. Rev. D \textbf{76}, 043513 (2007) [arXiv:hep-th/0702115].

\bibitem{Smolin1992} L. Smolin, \emph{Did the universe evolve?}, Class. Quantum Grav. \textbf{9}, 173 (1992).

\bibitem{BanksFischler2024} T. Banks and W. Fischler, \emph{Holographic inflation, primordial black holes and early structure formation}, Int. J. Mod. Phys. D \textbf{33} (2024) doi:10.1142/S0218271824400017 [arXiv:2402.11527].

\bibitem{Jacobson1995} T. Jacobson, \emph{Thermodynamics of spacetime: the Einstein equation of state}, Phys. Rev. Lett. \textbf{75}, 1260 (1995).

\bibitem{ChircoLiberati2010} G. Chirco and S. Liberati, \emph{Nonequilibrium thermodynamics of spacetime: the role of gravitational dissipation}, Phys. Rev. D \textbf{81}, 024016 (2010) [arXiv:0909.4194].

\bibitem{Bianconi2025} G. Bianconi, \emph{Gravity from entropy}, Phys. Rev. D \textbf{111}, 066001 (2025) [arXiv:2408.14391].

\bibitem{Bianconi2026} G. Bianconi, \emph{Thermodynamics of the gravity from entropy theory}, Phys. Rev. D \textbf{114}, 024042 (2026) [arXiv:2510.22545].

\bibitem{DorauMuch2026} P. Dorau and A. Much, \emph{Quantum relative entropy implies the semiclassical Einstein equations}, Phys. Rev. Lett. \textbf{136}, 091602 (2026).

\bibitem{BianconiEntropy2025} G. Bianconi, \emph{The quantum relative entropy of the Schwarzschild black hole and the area law}, Entropy \textbf{27}, 266 (2025).

\bibitem{CET2013} T. Clifton, G.~F.~R. Ellis and R. Tavakol, \emph{A gravitational entropy proposal}, Class. Quantum Grav. \textbf{30}, 125009 (2013).

\end{thebibliography}
\end{document}